# On the similarity of Information Energy to Dark Energy

M.P.Gough, T.D.Carozzi, A.M.Buckley

*Space Science Centre, University of Sussex, Brighton, BN1 9QT,UK*

m.p.gough@sussex.ac.uk, t.carozzi@sussex.ac.uk, a.m.buckley@sussex.ac.uk

**Abstract.** Information energy is shown here to have properties similar to those of dark energy. The energy associated with each information bit of the universe is found to be defined identically to the characteristic energy of a cosmological constant. Two independent methods are used to estimate the universe information content of ~$10^{91}$ bits, a value that provides an information energy total comparable to that of the dark energy. Information energy is also found to have a significantly negative equation of state parameter, $w < -0.4$, and thus exerts a negative pressure, similar to dark energy.

## 1 Introduction.

Information is directly bound up with the fundamental physics of nature. Some even consider the universe as being primarily comprised of information, summarized [1] by John Wheeler's slogan "it from bit". In any system of the universe the minimum unit of information, or bit, has associated with it a characteristic energy, $k_B T$, where $k_B$ is Boltzmann's constant and $T$ is the system temperature [2-8]. This information energy remains unavailable for doing work, effectively 'dark', except in cases where information is 'lost' from a system. One obvious example of information apparent non-conservation is the irreversible operation of standard digital logic where a full knowledge of the inputs can not be deduced uniquely from the output values [5-8]. Note that information is actually conserved with the information 'lost' still present in the nature of the dissipated energy. Special reversible logic circuits will eventually be necessary [2] [5] to prevent the small amount of information energy dissipated in digital logic from limiting available processing power as the energy necessary to store bits falls with shrinking electronics size. In a classical system description [3] the information content, $N$, of a system is given by $\log_2(m)$ bits, where $m$ is the number of degrees of freedom, or quantum states, present in that system.

A fundamental principle proposed [9] for quantum mechanics is that every particle or atom has many attributes (charge, mass, spin, etc, and aspects of position, direction and velocity, etc) that can each be considered as an elemental system with information content of one bit (or qubit). The quantum mechanical 'qubit' is synonymous [9] with the classical 'bit' in that both represent the most elementary unit of information. For this reason we hereafter refer only to the 'bit' and, for the purposes of our analogy, choose to emphasize the role of information instead of employing the related concept of entropy more generally considered in cosmology.

Here we show that the information energy in the universe has properties similar to dark energy with good agreement on the key parameters: characteristic energy; total energy; and equation of state parameter, *w*.

## 2 Characteristic energy.

In estimating the information energy of the universe it is clear that the various components of the universe each have a different representative temperature and hence a different characteristic energy associated with their information bits. Before considering such temperature differences it is instructive to first derive a value for the typical information bit energy in the universe considering the universe as a single component system, at one temperature, $T_u$. Then a sufficiently representative single temperature for the universe is obtained by equating $T_u$ to the temperature of a radiation dominated universe that has the same energy density as the matter of our universe,

$$\rho_{tot} c^2 = \alpha\, T_u^4 \qquad (1)$$

where $\rho_{tot}$ is the total matter density (including dark matter), and the radiation constant $\alpha = 4\sigma/c$. Describing the Stefan-Boltzmann constant, $\sigma$, in terms of fundamental constants, we obtain the information bit energy,

$$e_u = k_B T_u = [15\rho_{tot}\hbar^3 c^5/\pi^2]^{1/4} \qquad (2)$$

We note that this equation for the energy associated with an information bit in the simple single temperature universe is identical to that derived [10] previously for the characteristic energy of a cosmological constant. Equation 2 is identical to the equation 17.14 of Peebles [10] used to estimate the 3meV characteristic energy. This ultra low 3meV characteristic energy of a cosmological constant had been considered difficult to explain as it is too small to relate to any interesting particle physics [10]. Here we can simply associate this low energy with the characteristic information bit energy, $k_B T_u$.

Now a 3meV value for $k_B T_u$ corresponds to a low average temperature of 35K. This low average value today is consistent with much of the radiation occurring as 2.7K cosmic microwave background, CMB, and much of the matter as, possibly cold, dark matter, cold gas and dust, with only a small 0.4% of the universe, 1.7% of matter, [11] having formed stars, and primarily located in their cores at typically ~2 x $10^7$ K. While it is difficult to quantify the information bit content $N_c$ and temperature $T_c$ for each component, $c$, it is nevertheless clear that stellar interiors with their higher temperatures and hence much higher bit energies, even at only 1.7%, will make a major contribution to the information energy total.

## 3 Universe information content and total energy

In order to estimate the total information energy of the universe we need to calculate the total universe information bit content. Previously the information content of the matter and radiation in the universe has been estimated at $N_u \sim 10^{90}$ bits registered in quantum fields [4][12]. There are also possible gravitational quantum state degrees of freedom that could take the total towards the absolute maximum limit of $10^{120}$ bits set by the surface area of the universe expressed in Planck units according to the Holographic principle [12] [13]. However, for compactness, we follow the assumptions adopted previously [12] and we restrict ourselves here to information associated with the universe that is accessible by observation: matter and radiation.

Now the value of $N_u \sim 10^{90}$ bits was derived [12] from $(t/t_p)^{3/4}$ where $t$ is the age of the universe and $t_p$ is the Planck time. Further inspection of the equations involved shows this to be directly equivalent to dividing the total universe mass energy by the information characteristic energy, $e_u$, of equation (2) above. Clearly it is necessary to provide an independent estimate for $N_u$ in order for us to estimate the total information energy and at the same time avoid a 'circular argument'. Therefore we provide two such independent estimates for $N_u$ below.

Our first approach is to assume CMB provides a 'snapshot' of the universe at decoupling enabling us to estimate $N_u$ at decoupling from the information content of CMB today, and then account for the changes in the rest of the universe information since that time. Each CMB photon observed today has traveled uninterrupted since decoupling from a point on its spherical surface of last scattering. Each individual CMB photon's information is primarily its position of last scattering, in principle resolvable down to the present photon wavelength, $\sim 10^{-3}$m. Then the number of degrees of freedom for this position of scattering is $2 \times 10^{59}$, or $2^{197}$, given by the number of unit areas of size $10^{-3} \times 10^{-3}$ m$^2$ that fit into the total last scattering surface area, $2 \times 10^{53}$ m$^2$ for a sphere whose radius equals the distance light travelled since decoupling. This information per CMB photon has been constant from the time of decoupling as both CMB wavelength and size of the universe were a thousand times smaller then and have since scaled together to maintain a constant number of degrees of freedom. This is consistent with the intrinsic CMB information content remaining fixed over this period when CMB suffered negligible further interactions. Each of the $\sim 4 \times 10^{87}$ CMB photons therefore represents $\sim 197$ bits of information for a total $N_u$ at decoupling of $\sim 8 \times 10^{89}$. Since that time the universe has expanded by a factor of $10^3$ in size, or $10^9$ in volume, corresponding to an increase in information content of the rest of the universe by a factor of $\log_2(10^9)$, or 30. This approach therefore provides a present $N_u$ estimate of $\sim 2.4 \times 10^{91}$ bits.

Another approach is to assume that $N_u$ is given by that information needed by a hypothetical computer running a fully realistic classical simulation of the universe. Now the universe contains $\sim 10^{79}$ nucleons with the total number of all elementary particles, including photons, gravitons, neutrinos, etc, nine orders of magnitude higher at around $10^{88}$. Such a simulation would require enough information to track the movement of each fundamental particle at a resolution sufficient to describe the outcome of its interactions. In order to describe position and velocity in three dimensions over the scale of the universe, $\sim 10^{26}$m, with resolution of the order of a nucleon dimension, $10^{-15}$ m, we require an accuracy of one part in $10^{41}$, or one part in $2^{136}$. Thus our hypothetical simulation requires 136 bits for each of six dimensions, a total of 816 bits for each of the $10^{88}$ particles to provide an estimated $N_u \sim 8 \times 10^{90}$ bits. Note that this estimation is not very sensitive to the resolution used, since, even if we had assumed the resolution of a weakly interacting massive particle, or WIMP, with a Compton wavelength of say $10^{-25}$ m, the number of bits required to describe each particle would only increase to 1,020 bits and still imply $N_u \sim 10^{91}$ bits. Note also that those particles bound together within a larger system, for example a stellar interior, could be described just relative to that system and therefore might be expected to require considerably fewer bits. However, over a typical stellar diameter $\sim 10^9$m, resolutions of $10^{-15}$ m and $10^{-25}$ m still require 480 bits and 678 bits per particle respectively, and therefore such a consideration has little overall effect on our estimates.

These two independent estimates of $N_u \sim 10^{91}$ bits provide an information energy equal to the high dark energy value, assuming the above simple, single temperature, model with an average of 3meV per bit. This value of $10^{91}$ bits, thus associated with dark energy, compares with the previous estimate of $10^{90}$ bits, derived from considerations of mass energy [4] [12], with the difference in the two estimates to be expected from the difference between dark energy and total mass energy values.

## 4  Equation of State parameter

A key test for our comparison between information and dark energies is to see how information energy density varies with the cosmic scale factor, $a$. Now the energy densities of universe components vary as $a^{-3(1+w)}$ where $w$ is the equation of state parameter, relating pressure to energy density for each component. While $w_m = 0$ for matter, and $w_r = +1/3$ for electromagnetic radiation, dark energy is found [14] to exert a negative pressure with $w_{de}$ in the range: $-1.2 < w_{de} < -0.8$. This range includes the precise value, $w = -1$, required for a cosmological constant, but more data is needed to exclude other possibilities [14].

In our initial single temperature description above we expect information bit density to vary as $a^{-3}$ with universe expansion, initially assuming the total universe information content is fixed. We also expect the information characteristic energy, $e_u$, to simply change as $a^{-0.75}$ for the above single temperature $T_u$ dependence on $\rho_{tot}^{1/4}$ from equation (2). Unfortunately such a very simple model provides an information energy density that varies as $a^{-3.75}$, corresponding to $w_i = +\frac{1}{4}$, a value clearly incompatible with the presently accepted upper limit for dark energy, $w_{de} < -0.8$.

A more realistic model for information energy should include a representative multi-component composition with their individual system temperatures. It should also account for the increase in information content, or entropy, with time, consistent with the second law of thermodynamics.

The universe at the time of decoupling ($3.7 \times 10^5$ years age) can be approximated by a single component at ~3,000K, but the universe today ($1.4 \times 10^{10}$ years age) has clearly evolved into many components of widely different temperatures. While much of the 23% matter component has cooled to lower temperatures, some 1.7% of matter now exists [11] in stellar interiors. As noted in a previous section, these higher temperature components of today make a disproportionate contribution to the total information energy because of their higher intrinsic energy per bit.

Assuming information bit content is distributed in proportion to matter, we can see how the present information energy total, the total $\sum N_c k_B T_c$, summed over all universe components, c, compares with the earlier single component at decoupling. Here we make a simple approximation by noting that, although only 1.7%, the hot stellar matter at $2 \times 10^7$K will provide the dominant term today and we therefore neglect other terms, especially the colder matter terms. We can compare this to the situation at decoupling approximated by a single content-temperature product of 100% matter at $3 \times 10^3$K. Since characteristic information energy is directly proportional to component temperature this corresponds to an increase by a factor of $1.13 \times 10^2$ in total universe information energy since the time of decoupling if the information bit content remained constant. An increase in total information bits, or entropy, is also expected over this period from the change in distribution of matter over time, from the second law of thermodynamics with the number of degrees of freedom increasing as $a^3$. Classically, total information bit content increases as $\log_2 (a^3)$, corresponding to a factor of 30 since decoupling when the universe was a thousand times smaller.

Thus our first attempt at a more realistic model provides both an increasing component weighted average energy per bit by a factor of $1.13 \times 10^2$ and an increasing bit number by a factor of at least 30 since decoupling. Both factors contribute to an increase in total information energy to provide an information energy density that falls less steeply than $a^{-3}$, and hence provide a significantly negative $w_i$. In just comparing the two times, the time of decoupling with the present, we find that the total information energy has increased by a combined factor of $3.4 \times 10^3$ while $a$ increased by one thousand. This corresponds to an information energy density that changes by a factor of $3.4 \times 10^{-6}$ over this period, or to an average gradient between these two time points of $a^{-1.8}$, for $w_i = -0.4$.

We can expect the rarer high energy astrophysical objects with their even higher system temperatures to possess similarly disproportionate information energies over those of the more common stellar systems so that their inclusion would lead to a further reduction in $w_i$. Most importantly, our simple approach has simply compared the present time with decoupling and does not take into account how star formation varied over time. Combined results from a number of measurement techniques [15] show that less than 0.007 of the present star formation had occurred at the universe age of $10^9$ years. This provides a steeper increase in total information energy with time and therefore our first simple estimate comparing the present time with that at decoupling is clearly an upper limit: $w_i < -0.4$.

From an entropy point of view information is thought to increase on a cosmological scale with the increasing order that results from gravitational attraction [16]. This is in the opposite sense to the laboratory scale where the effects of gravitational attraction are negligible and entropy, and thus information, generally increases with increasing disorder. The information energy approach adopted here results in a universe information energy that increases with time due to the increase in average temperature caused by stellar formation. This increase is linear with temperature and therefore much greater than the reduction of information bit content which varies only logarithmically with the lost degrees of freedom on star formation. In section 3 we saw that the reduction in bits per particle on star formation was typically ~ 35% compared to the orders of magnitude increase in average temperature.

## 5 Summary

The main properties of information energy are similar in value to those of dark energy. Information has a characteristic energy identically defined to that of a cosmological constant and an information energy total comparable to the present high dark energy value. The evolution of higher temperature astrophysical objects causes the total information energy to increase over time and thus provide a negatively valued equation of state parameter. Our simply derived estimate for the information equation of state, $w_i < -0.4$ is consistent with the present range of estimates for the dark energy equation of state: $-1.2 < w_{de} < -0.8$. Thus information energy, like dark energy, exerts a negative pressure, in contrast to the $w \geq 0$ values of the other known universe components. We therefore conclude that information energy makes a significant contribution to dark energy.